\documentclass[mathleft
]{an}
\usepackage{graphicx}
\usepackage{times}
\usepackage{hyperref}
\begin{document}

\Yearpublication{2017}%
\Yearsubmission{2016}%
\Month{3}%
\Volume{-}%
\Issue{-}%
\DOI{10.1002/asna.201713334}%

\title{A multi-observatory database of X-ray pulsars in the Magellanic Clouds}

\author{J. Yang\inst{1,2,3}\fnmsep\thanks{Corresponding author:
  \email{jun.yang@cfa.harvard.edu}\newline}
S. G. T. Laycock\inst{1,2}, J. J. Drake\inst{3}, M. J. Coe\inst{4},  S. Fingerman\inst{1,2}, J. Hong\inst{3},  V. Antoniou\inst{3}, \and A. Zezas\inst{3}
}
\titlerunning{}
\authorrunning{J. Yang et al} 
\institute{
Lowell Center for Space Science and Technology, University of Massachusetts, Lowell, MA 01854.
\and 
Department of Physics and Applied Physics, University of Massachusetts,
    Lowell, MA 01854.
\and 
Harvard-Smithsonian Center for Astrophysics, Cambridge, MA 02138. 
    \and
Physics \& Astronomy, University of Southampton,
    SO17 1BJ, UK.
}

\publonline{8 March 2017}
\accepted{27 November 2016}
\received{30 September 2016}


\abstract{%
Using hundreds of XMM-Newton and Chandra archival observations and nearly a thousand RXTE observations, we have generated a comprehensive library of the known pulsars in the Small and Large Magellanic Clouds (SMC, LMC). The pulsars are detected multiple times across the full parameter spaces of X-ray luminosity ($L_X= 10^{31-38}$~erg/s) and spin period ( P$<$1s -- P$>$1000s) and the library enables time-domain studies at a range of energy scales. The high time-resolution and sensitivity of the EPIC cameras are complemented by the angular resolution of Chandra and the regular monitoring of RXTE. Our processing 
pipeline uses the latest calibration files and software to generate a suite of useful products for each pulsar detection: event lists, high time-resolution light curves, periodograms, spectra, and complete histories of $\dot{P}$, the pulsed fraction, etc., in the broad (0.2-12 keV), soft (0.2-2 keV), and hard (2-12 keV) energy bands. After combining the observations from these telescopes, we found that 28 pulsars show long-term spin up and 25 long-term spin down. We also used the faintest and brightest sources to map out the lower and upper boundaries of accretion-powered X-ray emission: the propeller line and the Eddington line, respectively. We are in the process of comparing the observed pulse profiles to geometric models of X-ray emission in order to constrain the physical parameters of the pulsars. Finally we are preparing a public release of the library so that it can be used by others in the astronomical community. }

\maketitle

\section{Introduction}
The Small Magellanic Cloud (SMC) is a dwarf irregular galaxy near the Milky Way at a distance of about 62 kpc (Graczyk et al. 2014; Scowcroft
et al. 2016). It contains a large and active population of X-ray binaries (e.g., Townsend et al. 2011, Haberl \& Sturm 2016).  The great majority (98\%) of the SMC high-mass X-ray binaries (HMXBs) are of the Be type (Coe et al. 2005). Be/X-ray binaries (Be-XBs) are stellar systems in which a Neutron Star (NS) accretes material from a massive early-type star with a circumstellar disc resulting in irregular bursts of energy. NSs are stellar remnants that can result from the gravitational collapse of massive stars after a supernova stage. They showcase extreme conditions, such as deep gravitational potential wells and strong magnetic fields, that are beyond the reach of laboratories here on Earth. 

Many more HMXBs are known in the SMC than the Large Magellanic Cloud (LMC) or the Milky Way (McBride et al 2008; Yokogawa et al 2003). Based on the relative masses of the Milky Way and the SMC, there is a factor of 50 more HMXBs than one would expect in the SMC (Walter et al 2006). The SMC has recently experienced an era of star formation, and the number of its HMXBs can be compared to the galactic star formation rate (Grimm et al 2003), which can provide us with a unique understanding of the end products of stellar evolution. 

The LMC at a distance of 50 kpc is 10 times more massive than the SMC. A possible explanation of its far fewer confirmed Be-XBs is the relation between the density of Be-XBs and the recent star formation activity in each galaxy. Antoniou et al. (2010) found that the ages of the stellar populations in which the Be-HMXBs are embedded are 25-60 Myr in the SMC compared to a younger population in the LMC (6-25 Myr old; Antoniou \& Zezas 2016). The LMC also features a wider range of HMXB types. 

\begin{figure*}
  \centering
 \includegraphics[width=1\linewidth]{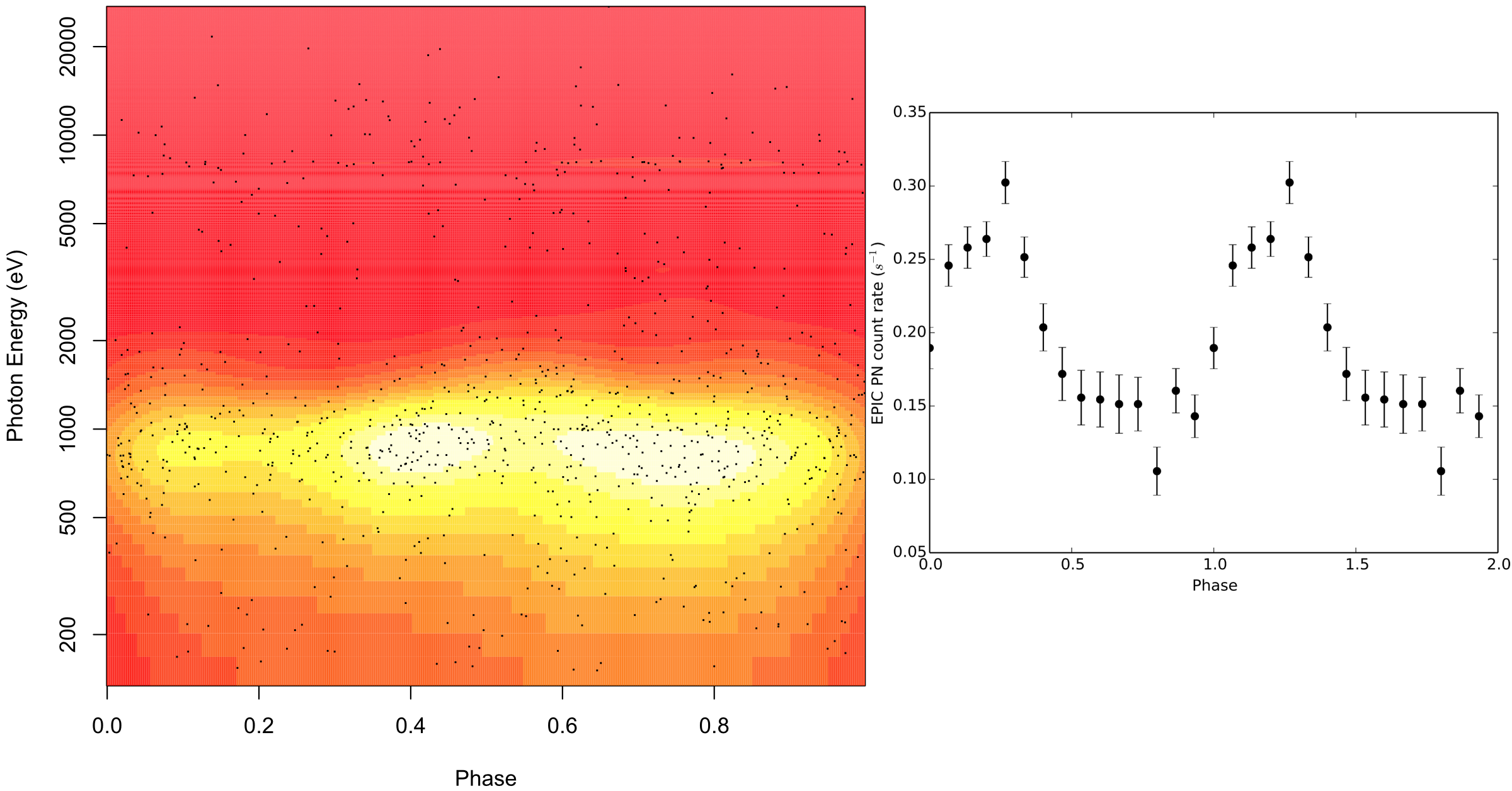}
\caption{\footnotesize{A phase-resolved pulsar event list (left: Photon Energy in eV $vs.$ Phase) and folded light curve (right: Photon Count Rate $vs.$ Phase) in the broad band (0.2 - 12 keV) for the X-ray pulsar SXP 504 from our library. Black dots in the left panel are the energy for each photon detected and the color contour shows the photon intensity.}}
\label{evt}
\end{figure*}

The low absorption and known distances of the Magellanic Clouds minimize uncertainties in the luminosities of the pulsars, both within the population and in absolute terms. This makes the Magellanic Clouds an ideal laboratory for understanding the process of accretion and X-ray emission in HMXBs during which the rapid rotation of a NS causes its radiating poles to \textquotedblleft{sweep}" over the Earth and reveal pulsations. The general picture of accretion onto X-ray pulsars consists of a flow in a wind or disk to the magnetosphere and then along the dipole field lines onto the magnetic poles of the NS. Although the basic picture has been known for decades, the details of accretion in places where the NS magnetic field dominates the flow is still a complex and challenging problem. As the accretion rate rises and X-ray luminosity increases, the observed beam pattern is often seen to change shape. The pulse profile morphology, energy dependence, and variations thereof are widely regarded as holding the key to developing a full physical picture of the accretion flow. It has been simulated in various theoretical approaches  for decades (e.g., Radhakrishnan \& Cooke 1969, Backer et al. 1976, Pechenick et al. 1983), but simple geometric models are still unable to fit many of the observed pulse profiles. 
Thus one of the goals of our project is to create such data products spanning the full range of behaviors in order to provide input to future more sophisticated pulse-profile models. 

Based on the surveys carried out by Chandra (e.g., Antoniou et al 2009; Laycock et al 2010), XMM-Newton (e.g., Haberl et al 2008), and RXTE (e.g., Laycock et al 2005, Klus et al 2014), we have generated a comprehensive time-domain library of the known Be-XB pulsars in the Magellanic Clouds, which contains all relevant publicly available observations performed by these satellites up to 2014. We have undertaken the relativistic modeling of the high-energy radiation and pulse profiles of the neutron stars in these systems to constrain the physics and geometry of the accretion flow. The paper is organized as follows: 
in \S~\ref{lib} we describe and discuss the overall properties of the library and some examples of extracted pulsar products. The underlying assumptions of our model and preliminary results of pulse-profile fitting are presented in \S~\ref{mod}. Finally, we outline in \S~\ref{sum} the summary and perspectives for future work.

\section{Examples of the library products}\label{lib}

Between 2000 and 2014, a total of 116 and 42 XMM-Newton observations of the SMC and LMC, respectively, were public. 
During the same period, Chandra observed the same targets for 155 and 30 times. We include these observations, as well as all RXTE monitoring of the Magellanic Clouds (which were targeted weekly between 1997 and 2012), in our analysis.

From the XMM-Newton archive, we have obtained the European Photon Imaging Camera (EPIC) PN data, which have a higher time resolution than the Metal Oxide Semiconductor (MOS) data, covering the energy range 0.1-15 keV. 
The standard procedures of the XMM-Newton Science Analysis Software (SAS, version 1.2)
were used for the data reduction.
Targets for analysis were from the SMC X-ray pulsar (SXP) catalog of Coe and Kirk (2015), which contains the spin period $P$ and celestial coordinates of each known pulsar. 
The LMC X-ray pulsar (LXP) list were chosen from \textquotedblleft{Research: Pulsars in the Magellanic Clouds}" catalog\footnote{Kept at \url{http://www.southampton.ac.uk/~mjcoe/}}.
The PN data were analyzed using standard commands like \textit{evselect} and \textit{epiclccorr} in SAS. Light curves and spectra of the pulsars were extracted from a circular region of radius 20$''$ and an annular background region (between radii of 50$''$ and 100$''$ from the source center) was used for background subtraction.
Similarly, we used ACIS data from Chandra and the reduction tools of the Chandra Interactive Analysis of Observations software package (CIAO, version 4.5; Fruscione et al. 2006). Source fluxes and light curves were extracted using \textit{srcflux} and \textit{dmextract}. 
The search for pulsations employed the Lomb-Scargle periodogram method (Press et al. 1992).

Altogether, our data products include single source event lists, pulse profiles, periodograms, spectra, and observed parameters such as X-ray flux, $P$, etc, from each detection.
The pulse profiles and periodograms were analyzed within the broad (0.2-12 keV), soft (0.2-2 keV) and hard (2-12 keV) energy bands. An example of an event list and folded light curve is shown in Figure~\ref{evt}. The left panel presents the photon energy as a function of phase from an XMM-Newton observation. Color contours show the photon intensities. We can see the highest photon intensity is located around 800 eV. The right panel shows the folded light curve the shape of which holds information about the emission patterns of the NS hot spots and the geometry of the source.

We also used the unique combination of capabilities of Chandra, XMM-Newton, and RXTE in order to obtain information about the long-term variations of the spin periods, X-ray luminosities, period amplitudes, and pulsed fractions of all sources and to measure their values of $\dot{P}$ over times that span at least 15 years. Some representative examples are presented below.

\subsection{X-ray Luminosity}

 \begin{figure}
 \centering
 \includegraphics[width=1\linewidth]{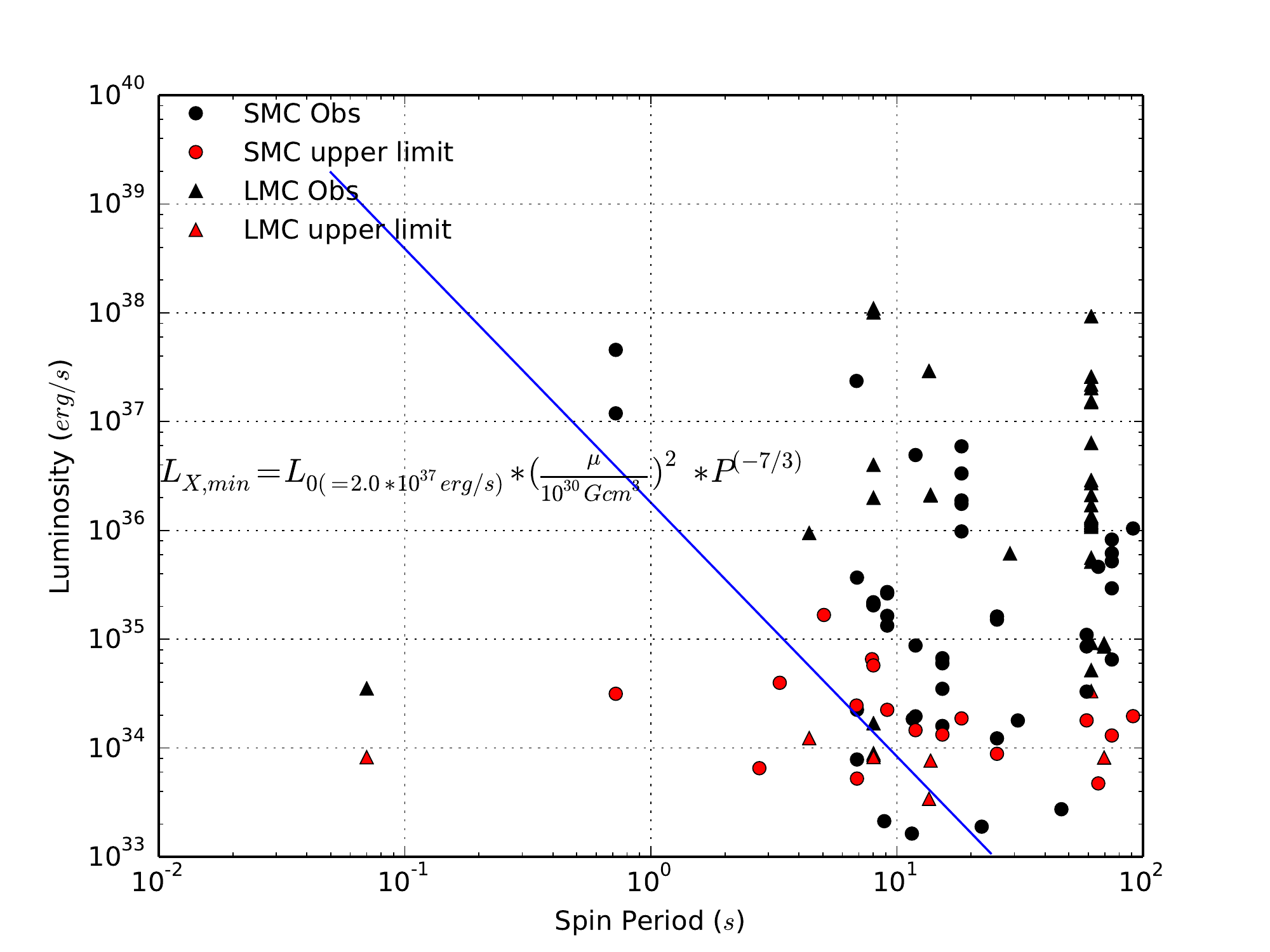}
\caption{\footnotesize{The $L_x$ $vs.$ $P$ diagram of X-ray pulsars in the SMC and LMC constructed from observations in the XMM-Newton science archive. Black symbols denote the XMM-Newton detections including both pulsed and un-pulsed sources and red symbols denote upper limits. The blue line is the theoretical P-line from Stella et al. (1986). Pulsations are seen only in detections above the P-line, although in this figure we do not have distinguished symbols for pulsed and un-pulsed pulsars.}}
\label{pline}
\end{figure}

\begin{figure*}
\centering
 \includegraphics[width=0.78\linewidth]{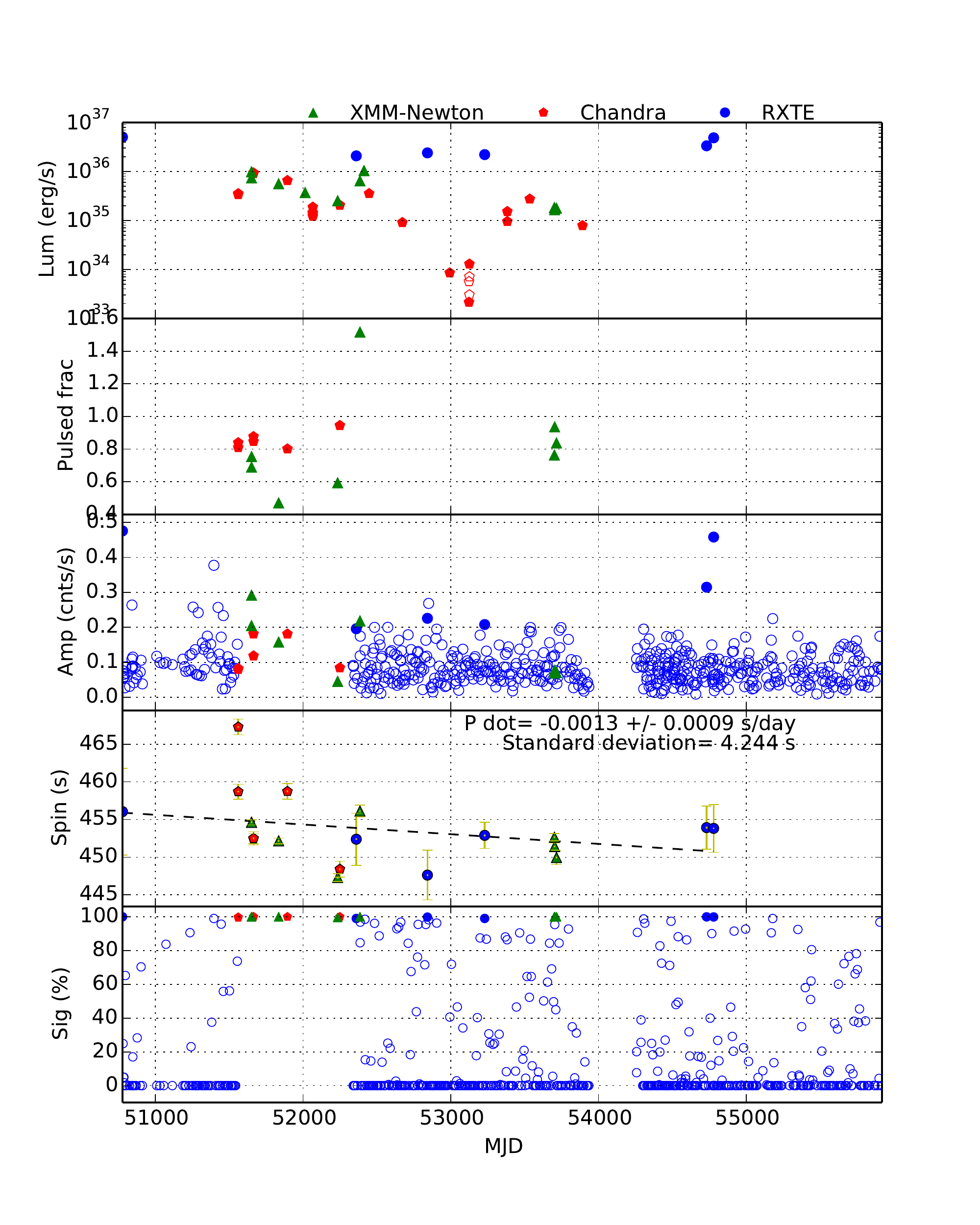}
  \caption{\footnotesize{Example of products for the pulsar SXP 455.  From top to bottom: luminosity, pulsed fraction, period amplitude, spin period, and significance of the Lomb-Scargle periods. Similar figures for 57 other pulsars will be presented by J. Yang et al. (2017).}}
  \label{pdot}
\end{figure*}

The accumulated X-ray luminosity ($L_X$) information of the known pulsars in the SMC and the LMC allows us to explore the relationships
between $L_X$ and other parameters such as spin period $P$. For example, Figure~\ref{pline} shows $L_x$  $vs.$ $P$ for the XMM-Newton observations of pulsars with $P < 100$~s. Upper limits to the X-ray luminosities were calculated from the \textquotedblleft{Flux Limits from Images from XMM-Newton}" (FLIX)\footnote{\url{http://www.ledas.ac.uk/flix/flix3.html}} server. The blue line is the propeller line (P-line; as presented by the formula in the figure). It is obtained from the equations given in Christodoulou et al (2016) and Stella \& White (1986) 
for a magnetic moment of $\mu = 3 \times 10^{29}$~G~cm$^3$, $M=1.4~M_{\odot}$ and $R=10~km$. No pulsations were detected in the few observations that landed below the P-line. These extremely faint detections could represent weak emission from the magnetospheres of the NSs. On the opposite end, some outbursts reach close to the Eddington limit that is located at $1.8\times 10^{38}$~erg~s$^{-1}$.

\subsection{Time evolution of the physical properties}
Figure~\ref{pdot} shows the time evolution of the X-ray luminosity, the pulsed fraction, the pulsed amplitude, $P$, and the significance of the Lomb-Scargle periods that we measured. Filled $vs.$ unfilled symbols describe the pulsar's
on/off status regarding pulsations. 
Chandra is more sensitive than XMM-Newton and therefore more appropriate to observe the source when it is in a low state. 
In the second panel, there is no pulsed fraction information for the non-imaging detector of RXTE since there were multiple un-resolved sources in the field of view (FOV). For epochs above Modified Julian Date (MJD) $\sim$52500 of Chandra observations, it is too faint to extract the light curve, so Chandra luminosities are shown in the first panel, but no information is shown in the other panels, as well as the XMM-Newton detection at MJD $\sim$52000. We do not calculate the upper limit of the RXTE since it is not reliable or useful.
In the third panel, unfilled circles indicate that the transient source is in the FOV, but no significant pulsations are detected to 99\% significance, which implies that the pulsar is in the quiescent state. In the fourth panel, we measure the period derivative, $\dot{P}$, by linear regression of all observations; for SXP 455, this is $-0.0013(9)$~s/day, so this pulsar is spinning up with the standard deviation of 4.244 s. Here the standard deviation is a measure of how spread out the spin periods around the best fitting slope are. 

\section{Geometric models of pulsar emission}\label{mod}
\begin{figure}
\centering
 \includegraphics[width=1\linewidth]{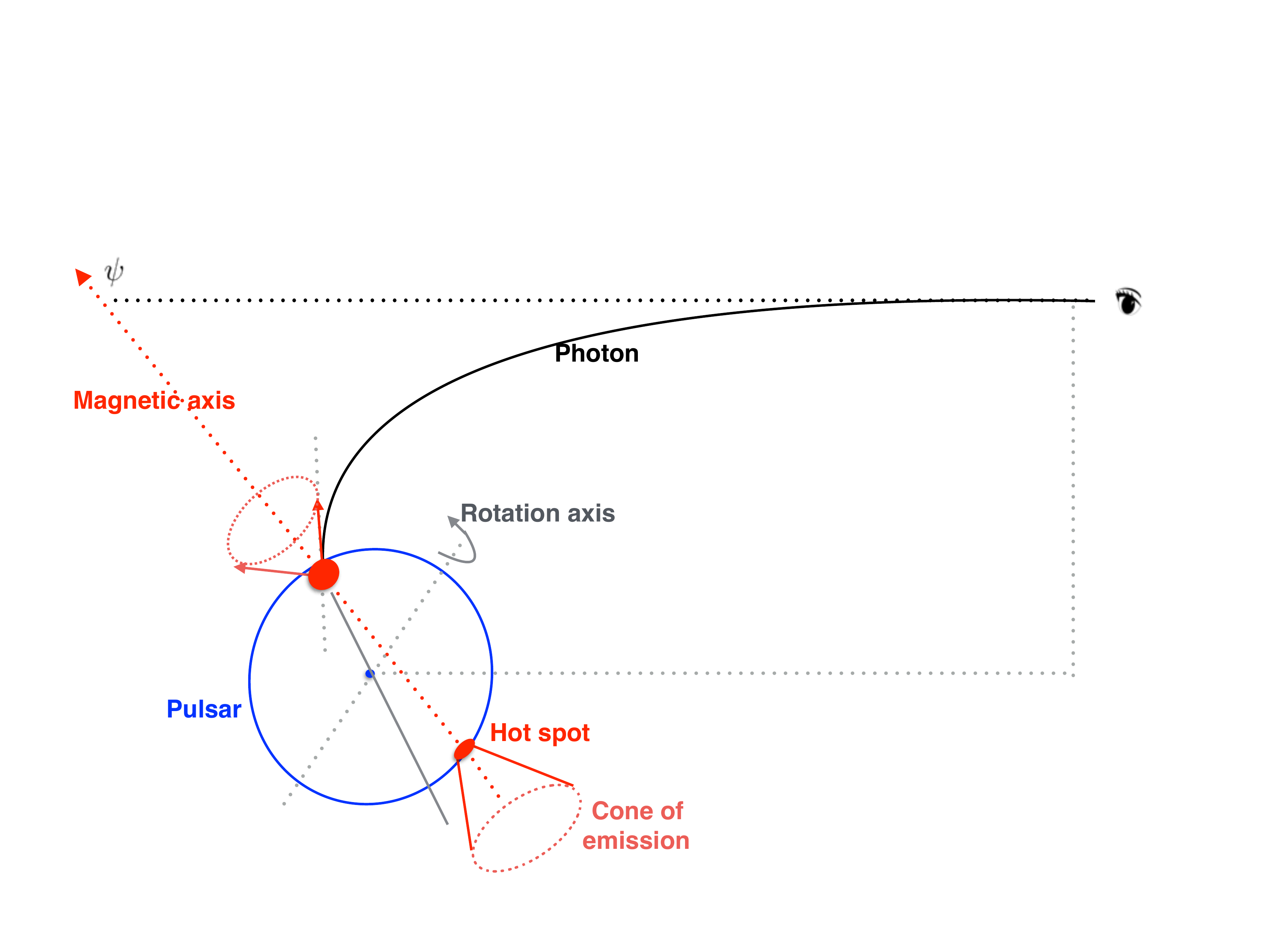}
  \caption{\footnotesize{ Illustration of an X-ray pulsar with an offset magnetic dipole axis.
The accretion hot spots are shown in red, while the
blue dot represents the center of the NS. The black line shows the
photon path to the observer including relativistic effects.}}
  \label{pulsar}
\end{figure}

Most of the existing literature assumes that NSs have two antipodal hot spots located at the magnetic poles and emitting isotropically (e.g., Beloborodov 2002). 
On the other hand, Backer et al. (1976) suggest an off-center magnetic axis with a displacement of 0.8-3~km from the stellar center.  
We have extended the model of Beloborodov (2002) to include an offset of the magnetic poles such that the magnetic axis does not have to pass through the center of the NS. A illustration of this geometry is shown in Figure~\ref{pulsar}. The red dotted line is the offset magnetic axis and the red ellipses represent the hot spots that can have an arbitrary size. The blue dot marks the center of the NS. The solid black line is the photon path including relativistic bending. 

\begin{figure}
\centering
 \includegraphics[width=1\linewidth]{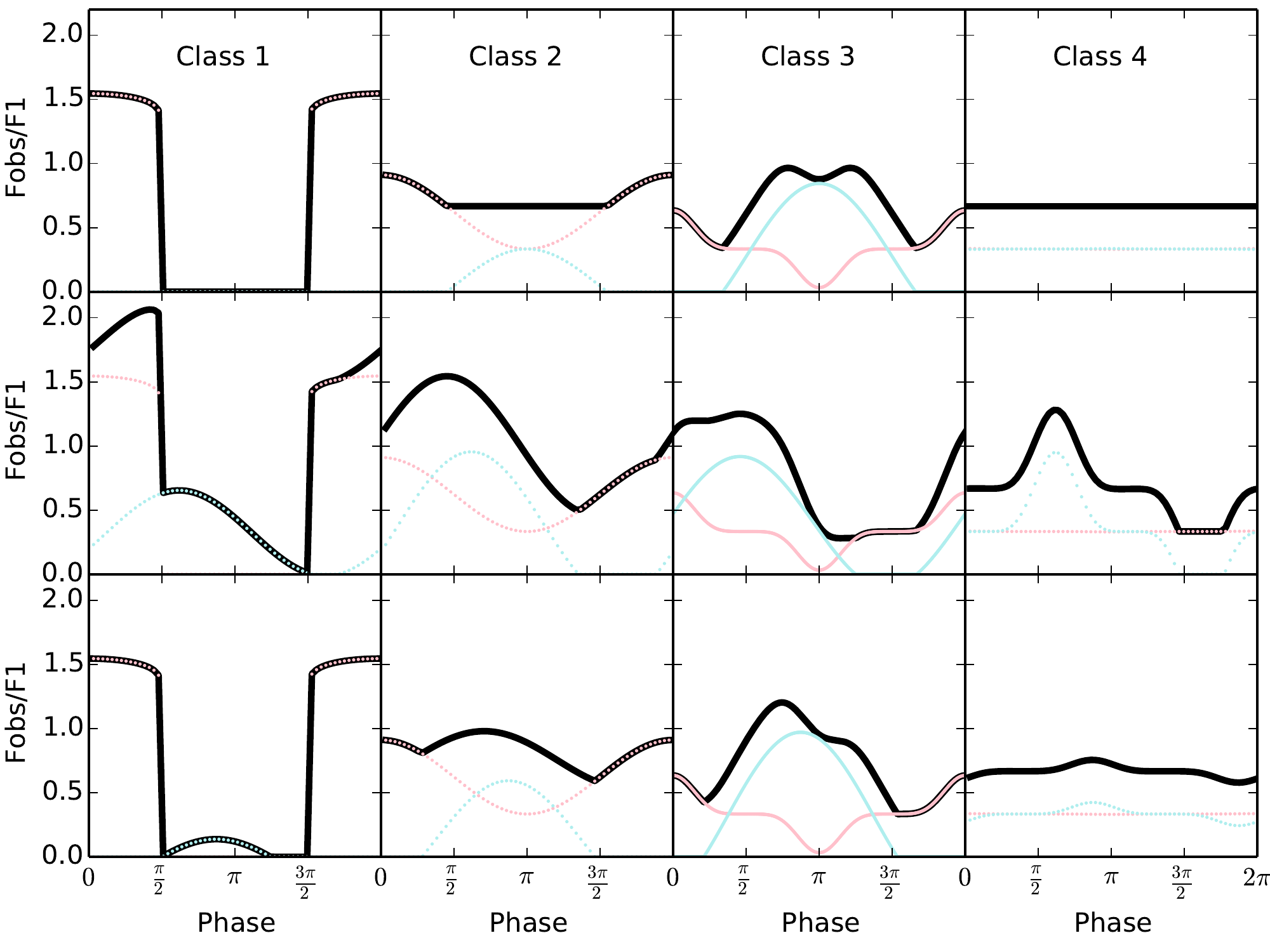}
  \caption{\footnotesize{Model light curves of the emission from the primary
hot spot (pink), secondary hot spot (cyan), and both hot spots combined
(black). Each class corresponds to a different angle combination between the magnetic axis, the rotation axis, and the line of sight.
Upper panels are for the pulsars with two antipodal hot spots and
no offset of the magnetic axis. Middle and bottom panels are for the
same pulsars except that the secondary hot spot is shifted in location.
(Fobs: observed flux, F1: maximum flux from one hot spot).}}
  \label{mod1}
\end{figure}

Based on this geometric model, we have created some pulse profiles of emission from an offset magnetic field (Figure~\ref{mod1}). The various classes are explained in the caption.
The observed flux from one hot spot is $F=F_1[\cos\psi (1-r_g/R)+r_g/R]$, where $\psi$ is the angle between the line of sight and the local radial direction of emission, $R/r_g$ is the emission radius in Schwarzschild units, and $F_1$ is the total flux emitted from this spot. At infinity, the flux is reduced by the gravitational redshift factor, $z_g$, 
which is given by $1+z_g=(1-r_g/R)^{-1/2}$. In classes 2 and 4 of the upper panel, and classes 3 and 4 of the middle panel of Figure~\ref{mod1}, when both hot spots are visible, the observed pulse shows a plateau $F_p=(2r_g/R)F_1$. The upper panels correspond to two antipodal hot spots. In this case, the maximum pulsed fraction of blackbody emission is $A_{max}=(R-2r_g)/(R+2r_g)$. Values $A > A_{max}$ can be produced by pulsars with asymmetric hot spots, as shown in the middle and bottom panels of Figure~\ref{mod1}. 

In this model, we can also vary the components of the beam pattern in order to make the pulse profile sharper or smoother. This parameterization changes the geometry of the intensities emerging from the hot spots which then needs to be propagated through the NS rotation and the  light bending.  Different model parameter combinations generate very different pulse profiles which can be compared to the observed profiles contained in our library. Our model appears to produce a larger spectrum of model profiles than the observations which implies that the observed pulsars may have certain similar properties that can be described by limited sets of parameters.  These aspects will be investigated in future work.

\begin{figure*}
\centering
 \includegraphics[width=1\linewidth]{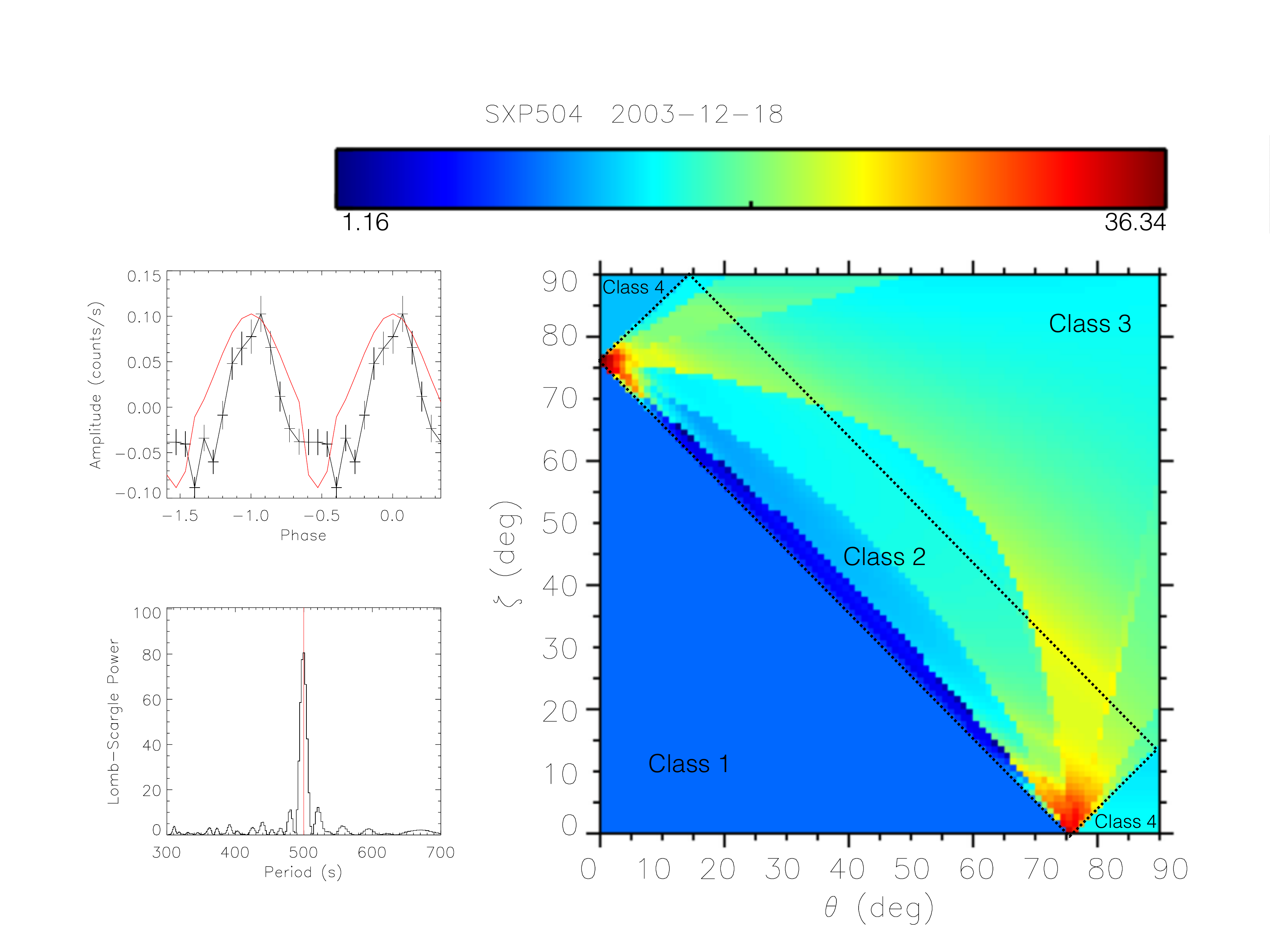}
  \caption{\footnotesize{Our preliminary pulse profile modeling for the pulsar SXP 504. 
  Top left: Folded light curve comparison between the 2003-12-18 XMM-Newton
observation (black) and best-fit model (red).
  Bottom left: Observed (black) and Lomb-Scargle model (red) periodograms.
  Right: Reduced $\chi^2$ map for the model input parameters (see text).}} 
  \label{fit}
\end{figure*}

A comparison between the observed pulse profile of source SXP 504 in the SMC and our geometric model is shown in Figure~\ref{fit}. The angle between the magnetic axis and the spin axis is $\theta$, while $\zeta$ is the angle between the line of sight and the spin axis; these are both free parameters. The color map shows the reduced $\chi^2$ between a 2003 XMM-Newton observation and the model at many equally-spaced grid points where it was computed. 
Although there has been no previous attempt to model this pulse profile, we obtained a minimum reduced $\chi^2$ of 1.16. 
The deduced best-fit parameters are $\theta=15^o$, $\zeta=64^o$, and $z_g=0.11$.

\section{Conclusion}\label{sum}
Using archival Chandra, XMM-Newton, and RXTE observations up to 2014, we have generated a comprehensive library of SMC/LMC X-ray pulsar observations which will be released for public use. Our time-domain library includes high time-resolution light curves, periodograms, spectra, event lists, and complete histories of $P$ and $\dot{P}$, pulsed fractions, and other physical quantities.  We are working toward extracting more physical properties from our library and incorporating more complex phenomena into our pulse-profile model. Pulse profile modeling is a very important tool in investigations of accretion flows and NS magnetospheres. The geometric model will be employed in fitting the known pulsars in the Magellanic Clouds. We anticipate that the resulting database will be of significant value to future theoretical research.  

As XMM-Newton enters its next decade of operation, it will play a crucial role in advancing the study of accretion-powered pulsars. The combination of sub-second photon-timing, CCD-grade energy resolution, and sensitivity to $L_x\sim10^{32}$ erg/s cannot be replicated by any other X-ray observatory. We suggest that XMM-Newton's role should be to expand the parameter space. 
The study of accreting X-ray pulsars employs the unique capabilities of the observatory and would benefit further from two suggested lines of action. Firstly, by providing increased-cadence monitoring of the Magellanic Clouds. We note that the weekly pointings of RXTE have presently provided most of the orbital periods and period derivatives, but RXTE is blind to the most interesting state transitions due to its low sensitivity and poor energy resolution. Secondly, 
by exploring the Local Group of galaxies, where ultra-luminous pulsars have been found (e.g., Bachetti et al. 2014). By means of long contiguous integrations (dozens of ksec), the unprecedented timing capabilities of XMM-Newton can unveil pulsations even if the angular resolution is not sufficient to separate individual sources. Other Local Group galaxies would represent new astrophysical laboratories for studying the role of the environment on the star formation history and on pulsar properties.

\acknowledgements
We acknowledge support from NASA-ADAP grant NNX14-AF77G. We thank the anonymous referee for the very helpful and constructive comments. JY thanks the American Astronomical Society International Travel Grant program for enabling her participation in the XMM-Newton Next Decade meeting. 


\end{document}